%% file: paper.tex
\documentclass[12pt]{article}
\usepackage{graphicx}
\usepackage{amsmath}

\newcommand\pubnumber{NuPhys2017-Leardini}
\newcommand\pubdate{\today}

\def\unipd{Universit\`a degli Studi di Padova, 35131 Padova, Italy}

\def\Title#1{\begin{center} {\Large #1 } \end{center}}
\def\Author#1{\begin{center}{ \sc #1} \end{center}}
\def\Address#1{\begin{center}{ \it #1} \end{center}}

\newcommand\pubblock{\rightline{\begin{tabular}{l} \pubnumber\\
         \pubdate  \end{tabular}}}
\newenvironment{Abstract}{\begin{quotation}  }{\end{quotation}}
\newenvironment{Presented}{\begin{quotation} \begin{center} 
             PRESENTED AT\end{center}\bigskip 
      \begin{center}\begin{large}}{\end{large}\end{center} \end{quotation}}
\def\Acknowledgements{\bigskip  \bigskip \begin{center} \begin{large}
             \bf ACKNOWLEDGEMENTS \end{large}\end{center}}

\input econfmacros.tex

\begin{document}
\begin{titlepage}
\pubblock

\vfill
\Title{A study about the F-estimator for the neutrino mass hierarchy in the JUNO experiment}
\vfill
\Author{ Sara Leardini}
\Address{\unipd}
\vfill
\begin{Abstract}
\noindent At present, it is still unknown whether the correct mass ordering of the neutrino mass eigenstates
is either $m_1$, $m_2$, $m_3$ (Normal Hierarchy, NH), or $m_3$, $m_1$, $m_2$ (Inverted Hierarchy, IH). 
The new analysis method proposed by Stanco et al.~\protect\cite{Stanco} should fix some issues of the currently most used estimator, 
$\Delta \chi^2$, and make it possible to reach $5\,\sigma$ measurements in less than
six years of data taking with JUNO (Jiangmen Underground Neutrino Observatory, ~\protect\cite{An}, ~\protect\cite{Ranucci}) if a degeneracy in the atmospheric mass, $\Delta m^2_{atm}$, is accepted. In this note, the analysis introduced in the paper above was extended to more detailed studies on the dependence of the new F estimator to $\Delta m^2_{atm}$. A fit to the values of the new estimator as a function
of $\Delta m^2_{atm}$, calculated for both the true hierarchy and the wrong hierarchy, was performed. \\
The study of the fitting function showed that the average minimum counts, corresponding to the best value for $\Delta m^2_{atm}$, either for the true hierarchy or for the wrong hierarchy, are well separated, and allow to distinguish easily between NH and IH.

\end{Abstract}
\vfill
\begin{Presented}
NuPhys2017, Prospects in Neutrino Physics\\
Barbican Centre, London, UK,  December 20--22, 2017
\end{Presented}
\vfill
\end{titlepage}
\def\thefootnote{\fnsymbol{footnote}}
\setcounter{footnote}{0}

\section{Introduction}
One of the most compelling issues in Neutrino Physics is the determination of the neutrino mass hierarchy. So far, for the three neutrino mass eigenstates, $\nu_1$, $\nu_2$ and $\nu_3$, only the following quantities have been measured: the "solar" mass term $\Delta m^2_{sol}=\Delta m^2_{12}=m^2_2-m^2_1$, and the abolute value of $|\Delta m^2_{31}| \simeq |\Delta m^2_{32}|$. As a consequence, two possible mass ordering are allowed: the normal hierarchy (NH), with $m_1^2 < m_2^2 < m_3^2$, and the inverted hierarchy (IH), with $m_3^2 < m_1^2 < m_2^2$. In this paper the notation "$\Delta m^2_{atm}$" (which indicates the atmospheric mass term) will be used, either to indicate $\Delta m^2_{31}=m^2_3-m^2_1$ in the case of normal hierarchy, or to indicate $\Delta m^2_{23}=m^2_2-m^2_3$ in the case of inverted hierarchy. \\ \\
\noindent For the studies regarding the mass hierarchy determination, only one estimator has been extensively used so far, the $\Delta \chi^2$ test:
\begin{align}  
\Delta\chi^2=\chi^2_{min}(IH)-\chi^2_{min}(NH)
\end{align}
\noindent where $=\chi^2_{min}(IH)$ and $\chi^2_{min}(NH)$ come from the best-fit values for IH and NH respectively. The fit is performed over the whole set of the uncertainty parameters, namely the neutrino oscillation ones and the systematic errors. Nevertheless, this estimator has caused some concerns ~\protect\cite{Ciuffoli}. 
\section{The F-estimator for reactor neutrinos}
Reactor anti-neutrinos with an energy $E_{\nu}$ can be detected in a scintillator at a distance $L$ from the reactor where they produced from the inverse $\beta$ decay
$\bar{\nu}_e + p \rightarrow e^+ + n$. The rate of detected events is given by
\begin{align}
\frac{dN}{dE_{\nu}}= T \times \Phi (E_{\nu}) \times \sigma_{\bar{\nu}_e p} \times P_{\bar{\nu}_e \rightarrow \bar{\nu}_e},
\end{align}

\noindent where $T$ is the thermal power of the reactor, $\Phi$ and $\sigma_{\nu_e p}$ are respectively the flux and the cross section
of the anti-neutrinos, and $P_{\nu_e \rightarrow \nu_e}$ is their survival probability. For every fixed value of the atmospheric mass, either equal or distinct for NH or IH, the survival probability is different for NH and IH. Its difference brings to the quantity $\Delta N(E_{\nu})=
\left(  \frac{dN}{dE_{\nu}}\right)_{NH}-\left(  \frac{dN}{dE_{\nu}}\right)_{IH}$, which is modulated at first order by $\sin (\frac{L}{2E}\Delta m^2_{atm})$. \\ 
By using this fact, Stanco et al. ~\protect\cite{Stanco} introduced a new estimator, called F-estimator, defined as
\begin{align}
F_{MO}=\int\limits_{1.8}^{8.0} |\Delta N(E_{\nu})|\,dE_{\nu}
\end{align}
\noindent where the limits of the integral are chosen to include the energy interval in which the modulation induced by $\Delta m^2_{atm}$ is observable. \\
In order to obtain the value of F, two kinds of intervals have to be defined:
\begin{enumerate}
 \item[1)]{$I^{+}=\{E_{\nu}: N_{NH,exp}(E_{\nu})>N_{IH,exp}(E_{\nu})\}$, i.e. the values of energy where the number of expected events for NH is greater 
 than the number of expected events for IH;}
 \item[2)]{$I^{-}=\{E_{\nu}: N_{NH,exp}(E_{\nu})<N_{IH,exp}(E_{\nu})\}$, i.e. the values of energy where the number of expected events for NH is less 
 than the number of expected events for IH;}
\end{enumerate}
\noindent F is then computed both for the normal and the inverted hierarchy in the following way:\\ \\
\noindent $F_{IH}=\int\limits_{1.8}^{8.0} (N_{obs}-N_{IH,exp}) \,dE_{\nu}$ in $I^{+}$ if $N_{obs}>N_{IH,exp}$ ; \\
$F_{IH}=\int\limits_{1.8}^{8.0} (N_{IH,exp}-N_{obs}) \,dE_{\nu}$ in $I^{-}$ if $N_{obs}<N_{IH,exp}$ ; \\
$F_{NH}=\int\limits_{1.8}^{8.0} (N_{obs}-N_{NH,exp}) \,dE_{\nu}$ in $I^{-}$ if $N_{obs}>N_{NH,exp}$ ; \\
$F_{NH}=\int\limits_{1.8}^{8.0} (N_{NH,exp}-N_{obs}) \,dE_{\nu}$ in $I^{+}$ if $N_{obs}<N_{NH,exp}$ . \\ \\
\noindent In the ideal case, if NH is true, $F_{NH}$
is equal to zero, and $F_{IH}\sim 6500$ in 6 years of JUNO data taking ~\cite{Stanco},~\cite{An}; vice versa if IH is true. In the real case, both
$F_{NH}$ and $F_{IH}$ are different from zero, yet they are different enough
to have a high sensitivity, if a long exposure is performed and a good energy resolution is obtained.
\section{The analysis}
\begin{figure}[htb]
\centering
\includegraphics[height=1.5in]{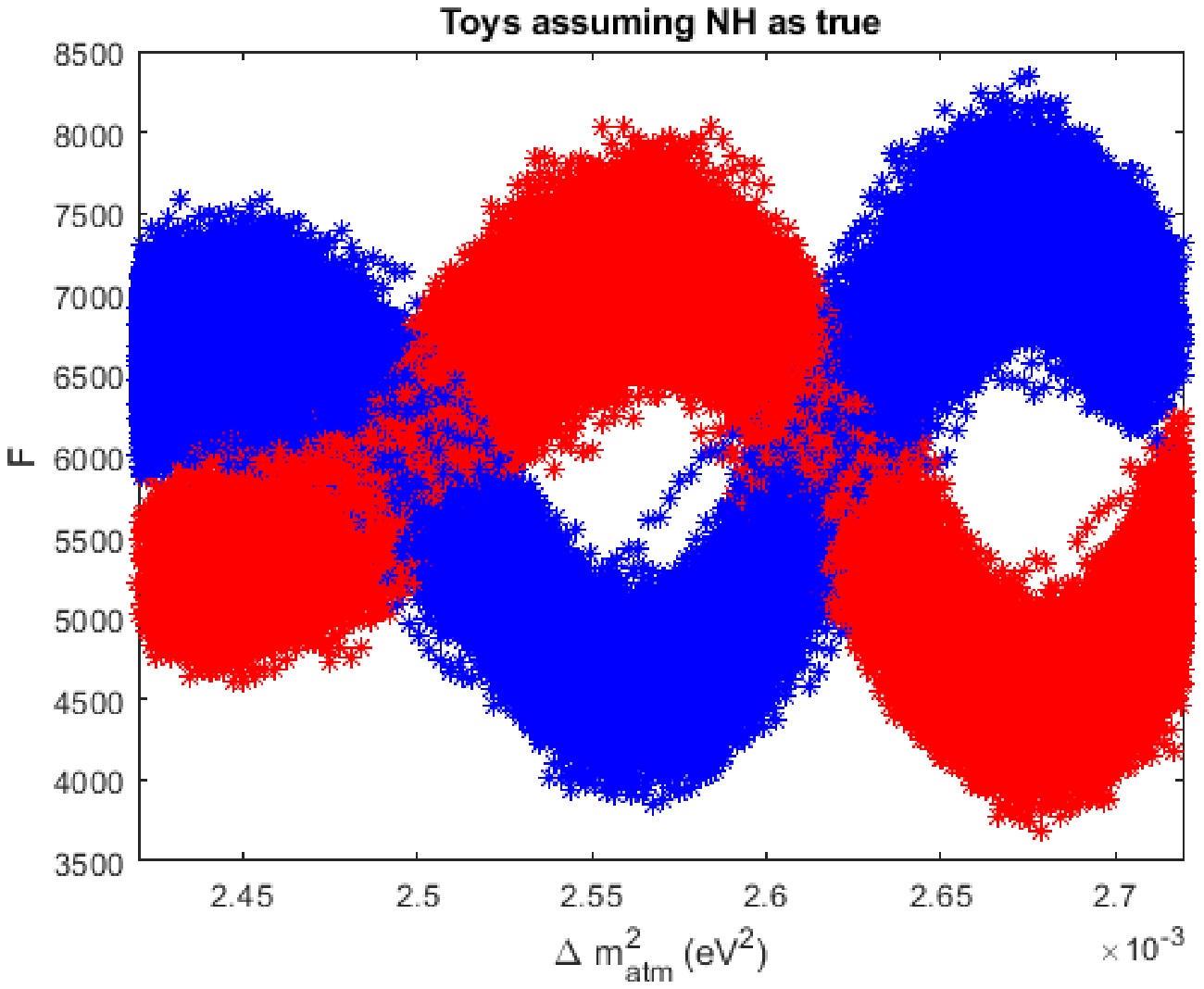}\,
\includegraphics[height=1.5in]{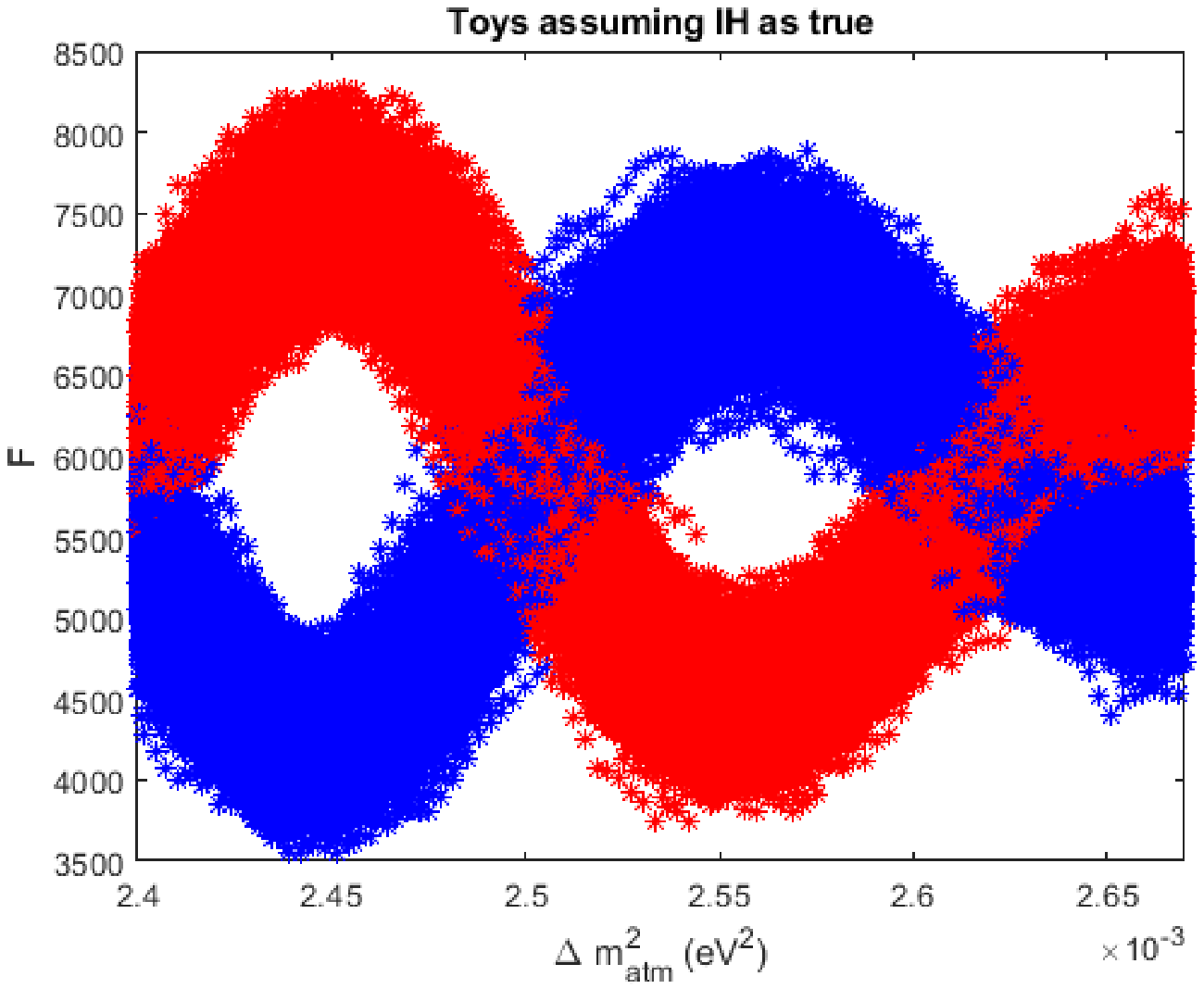}
\caption{Plots of F vs $\Delta m^2_{atm}$ for a universe in which NH is true (left) and a
universe in which IH is true (right). The strip-like form is due to the fact that all the 1000 toys are superimposed; the values of F for NH are blue, the values
of F for IH are red.}
\label{fig:toys}
\end{figure}

\begin{figure}[htb]
\centering
\includegraphics[height=1.5in]{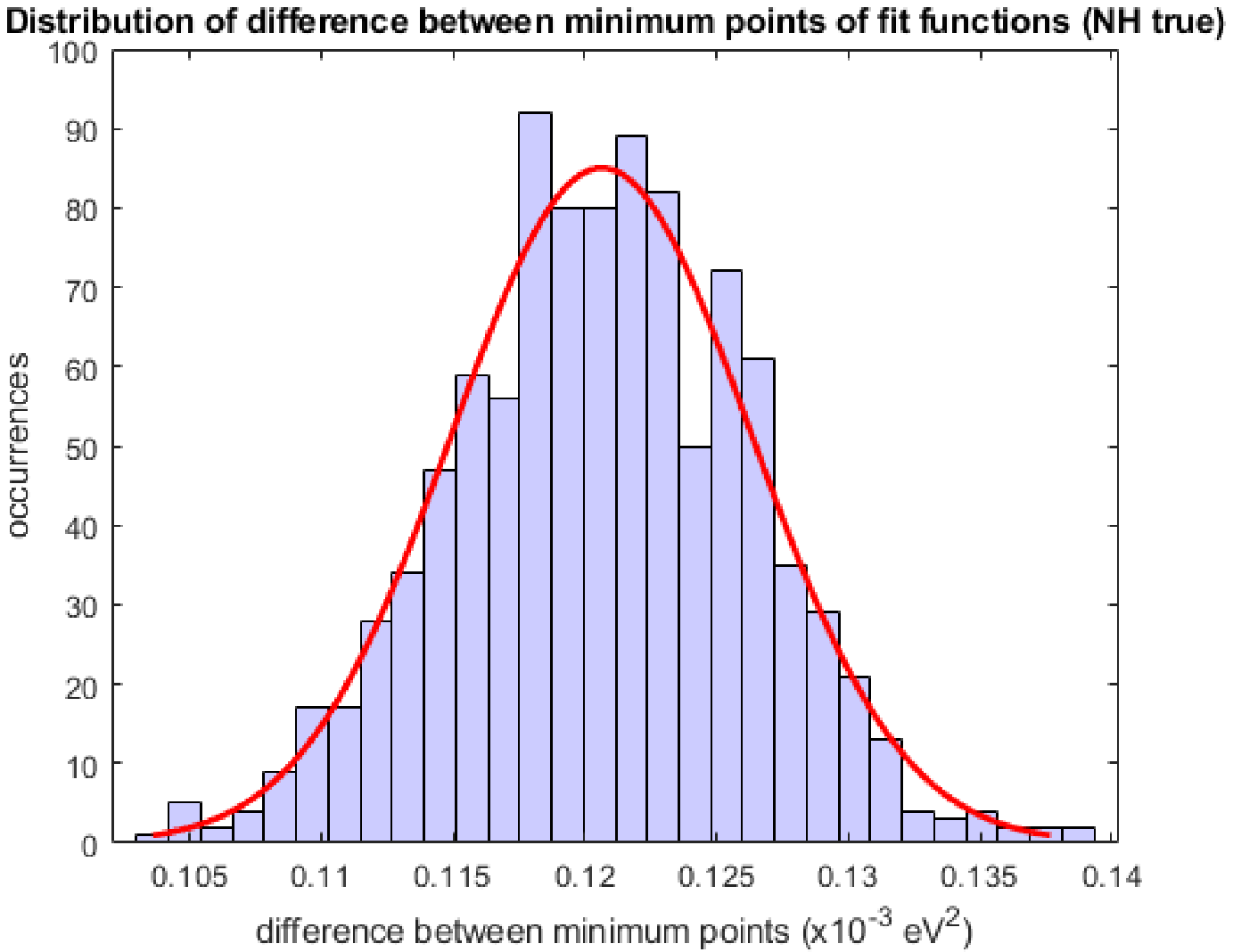}\,
\includegraphics[height=1.5in]{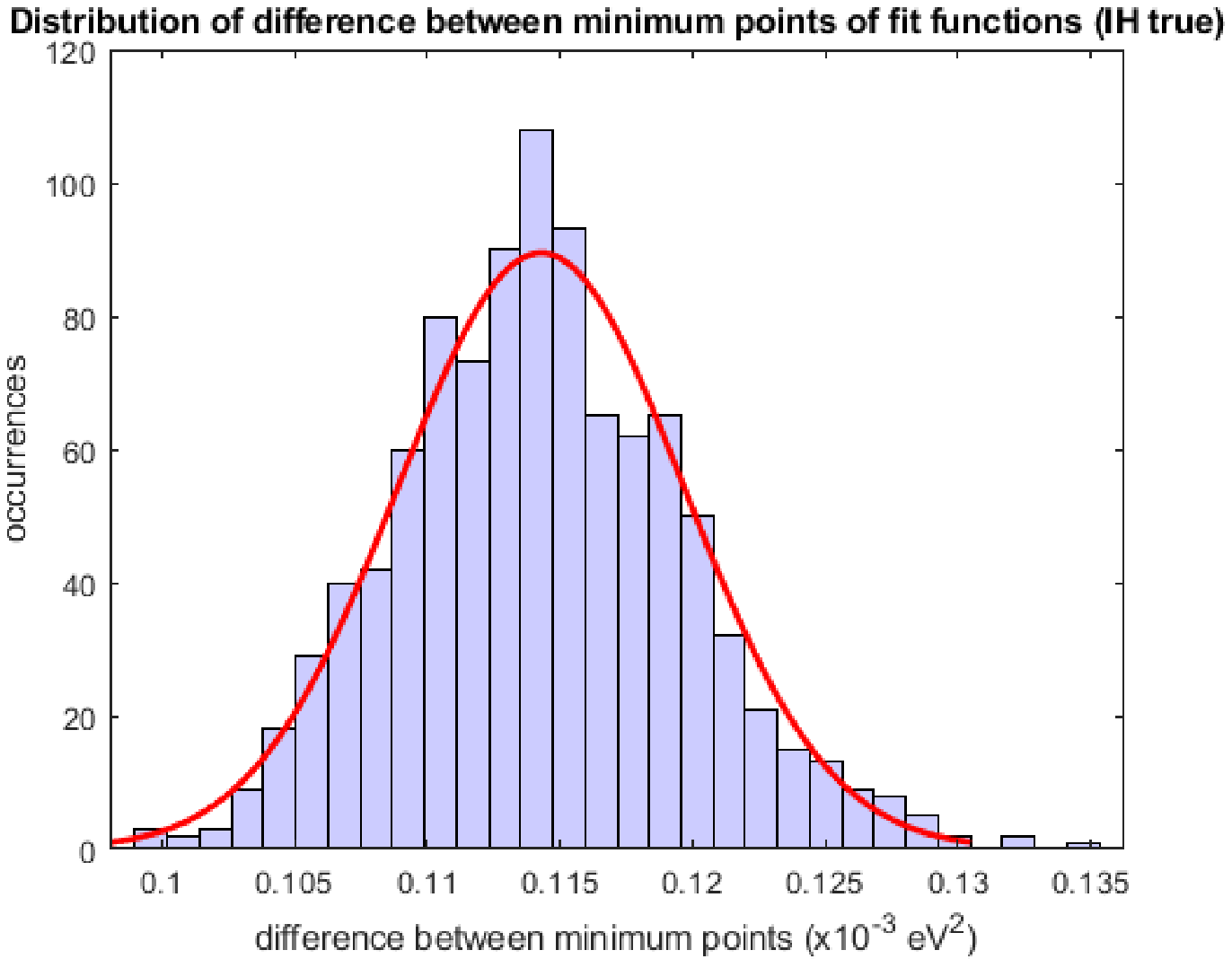}
\caption{Histograms showing the distributions of $\Delta min$ for NH (left) and IH (right).}
\label{fig:minima}
\end{figure}

The analysis presented in this paper makes use of 2000 toys, generated by Monte Carlo simulations of JUNO-like events
(1000 toys assuming NH and 1000 toys assuming IH). $\Delta m^2_{atm}$ was set equal to $2.56\cdot 10^{-3}$ eV$^2$, for both NH and IH, at the generation level. 
Further, for each toy F$_{NH}$ and F$_{IH}$ were computed assuming 101 different values of $\Delta m^2_{atm}$. In this way it could be checked whether the values of $\Delta m^2_{atm}$ for which F$_{NH}$ and F$_{IH}$ present a minimum correspond to the true minimum and are distinguishable with a good level of confidence.
In Figure \protect\ref{fig:toys} the plots of F vs $\Delta m^2_{atm}$ are shown, both for the universe with NH and the one
with IH. For every toy a fit to F$_{NH}$ and F$_{IH}$ with the function $\boldmath{F_{fit}(\Delta m_{atm})=A\cos(\omega_1 \Delta m_{atm}+\Phi_1)\cos(\omega_2 \Delta m_{atm}+\Phi_2)+h}$ 
was performed. $A$, $\omega_1$, $\omega_2$, $\Phi_1$, $\Phi_2$ and $h$ are the free parameters for the fit.
\noindent The positions of minimum min$_{true(false)}$ and maximum max$_{true(false)}$ for the fitting function were evaluated for every toy. Before any further probe, the values obtained were bias corrected. In fact, because of the presence of a bias due to the energy resolution that depends on the energy itself, F was shifted by some quantity ~\protect\cite{Stanco}. Hence, it was necessary to restore it to the right position, such that the minimum min$_{true}$ (maximum max$_{false}$) corresponded to $\Delta m_{atm}^2=2.56\cdot 10^{-3}$ eV$^2$. The position of the minima for the wrong hierarchy was then determined, and its distance from the true value, 2.56 $\cdot 10^{-3}$ eV$^2$, was evaluated for every toy: $\Delta $min=$|$min$_{false}$-min$_{true}|$. \\
\noindent The values obtained for both NH true and IH true are (see also Figure \protect\ref{fig:minima}): \\ \\
\noindent $<\Delta $min$>=(12.06\pm 0.02 )\times 10^{-5}$, $\sigma_{\Delta min} =(0.57 \pm 0.01)\times 10^{-5}$ for NH \\
$<\Delta $min$>=(11.43\pm 0.02)\times 10^{-5} $, $\sigma_{\Delta min} =(0.54 \pm 0.01)\times 10^{-5}$ for IH,  \\
\noindent the $\sigma_{\Delta min}$s being the dispersions of the false minima.

\noindent These results are perfectly consistent with the suggested value of $12\times 10^{-5}$ eV$^2$ ~\cite{Stanco}. 
This allows to discriminate between the two hierarchies if the value of $\Delta m_{atm}^{2}$ is known with a precision of $\sim 4$\%. By comparison, the $\Delta \chi^2$ test requires a precision of $\sim 2$ \% on the $\Delta m_{atm}^{2}$ value.





\section{Conclusions}
From the analysis shown in this work, the new estimator introduced in ~\protect\cite{Stanco} seems to provide promising results. In particular, the precision required on $\Delta m^{2}$ results to be more relaxed than the precision needed for the $\Delta \chi^2$ test. \\ Even if not mentioned here, an even more relevant feature of F is the increase of its significance with regards to the collected luminosity (see ~\protect\cite{Stanco}). In contrast, the $\chi^2$ tends asymptotically to a limited significance. We have further studied the characteristics of the F estimator, as described in this note. 
By focusing on the precision on $\Delta m^2_{atm}$ with which F estimates the best value, either for NH or IH, we were able to conclude that an excellent precision is achievable, confirming the rough estimation done in ~\protect\cite{Stanco}. \\

\Acknowledgements
I would like to thank the organisers of "NuPhys2017" Dr. Peter Ballett, Prof. Francesca Di Lodovico and Prof. Silvia Pascoli for arranging such interesting and stimulating conference, and for accepting my poster. I am also grateful to Prof. Luca Stanco for supervising my work, for providing me precious explanations and for having enough trust in me to let me present a poster, despite I am still an undergraduate student.

\end{document}

%% file: econfmacros.tex
\def\beq{\begin{equation}}
\def\eeq#1{\label{#1}\end{equation}}
\def\eeqn{\end{equation}}

\def\beqa{\begin{eqnarray}}
\def\eeqa#1{\label{#1}\end{eqnarray}}
\def\eeqan{\end{eqnarray}}

\let\bar=\overbar

\def\Dslash{\not{\hbox{\kern-4pt $D$}}}
\def\dslash{\not{\hbox{\kern-2pt $\del$}}}

\def\msb{{\bar{\ssstyle M \kern -1pt S}}}

